\documentstyle[ltwol]{article}

\def\be{\begin{equation}}
\def\ee{\end{equation}}
\def\bea{\begin{eqnarray}}
\def\eea{\end{eqnarray}}

\begin{document}

\title{RECENT PROGRESS OF LATTICE QCD IN CHINA}

\author{Xiang-Qian LUO, Eric B. GREGORY}

\address{Department of Physics, Zhongshan University, Guangzhou
510275, People's Republic of China\\E-mail: stslxq@zsu.edu.cn, 
steric@zsu.edu.cn}   

\twocolumn[\maketitle\abstracts{Lattice QCD 
is the most reliable non-perturbative method
in quantum field theory. In the last few years, some problems crucial to high energy experiments 
have been solved.
We review some recent work done by the Chinese lattice
community.}]

\section{Introduction}

Particle physics and gauge field theories are frontiers of fundamental sciences of matter. 
According to Yang and Mills, any basic theory of interacting matter should satisfy gauge invariance. 
QCD has been accepted to be the most successful gauge theory of strongly interacting particles. 
In QCD, hadronic matter is composed of quarks, and interactions between them are mediated 
by eight massless gluons generated by the SU(3) gauge group. 
Asymptotic freedom of QCD at short distances makes the perturbative calculations 
of high-energy processes possible. 
At low energy scales, however, QCD has many important properties which can not be studied 
perturbatively,
like confinement of quarks and gluons, 
vacuum structure, chiral-symmetry breaking, glueball masses, hadronic spectrum 
and weak interaction processes, 
and behaviors of hadronic matters at high temperature or high density.

Lattice gauge theory (LGT) 
has developed into a promising first principle 
non-perturbative approach to these phenomena. 
The basic idea \cite{Wilson:1974sk}, as proposed by K. Wilson in 1974, 
is to replace the continuous space and time 
by a discrete grid:

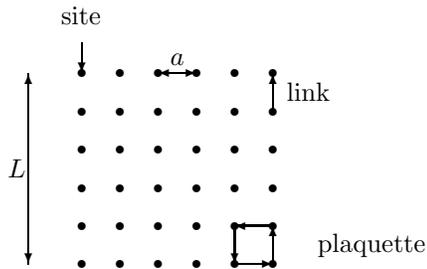
\begin{center}
\setlength{\unitlength}{.02in}
\begin{picture}(80,80)(0,0)

\multiput(25,5)(10,0){6}{\circle*{2}}
\multiput(25,15)(10,0){6}{\circle*{2}}
\multiput(25,25)(10,0){6}{\circle*{2}}
\multiput(25,35)(10,0){6}{\circle*{2}}
\multiput(25,45)(10,0){6}{\circle*{2}}
\multiput(25,55)(10,0){6}{\circle*{2}}

\put(11,30){\vector(0,2){25}}
\put(11,30){\vector(0,-2){25}}
\put(8,30){\makebox(0,0){$L$}}

\put(50,55){\vector(1,0){5}}
\put(50,55){\vector(-1,0){5}}
\put(50,60){\makebox(0,0)[t]{$a$}}

\put(25,63){\vector(0,-1){8}}
\put(30,70){\makebox(0,0)[r]{site}}

\put(75,45){\vector(0,1){10}}
\put(90,50){\makebox(0,0)[r]{link}}
\put(75,45){\line(0,1){10}}

\put(65,5){\vector(1,0){10}}
\put(75,5){\vector(0,1){10}}
\put(75,15){\vector(-1,0){10}}
\put(65,15){\vector(0,-1){10}}
\put(115,10){\makebox(0,0)[r]{plaquette}}
\end{picture}
\end{center}

\vskip 0.5cm

\noindent
Gluons live on links $U(x,\mu)=e^{ig\int_x^{x+ \hat{\mu}a} dx' A_{\mu}(x')}$, and quarks 
live on sites.  
The continuum Yang-Mills action $S_g=\int d^4x ~\rm{Tr} F_{\mu \nu}(x)F_{\mu \nu}(x)/2$ is replaced by
\begin{eqnarray}
S_g={1 \over g^2} \sum_p \rm{Tr} (U_{p} +U_{p}^{\dagger}-2),
\label{gauge}
\end{eqnarray}
where $U_p$ is the ordered product of link variables $U$ around an elementary plaquette.
The continuum quark action $S_q=\int d^4x ~\bar{\psi} (x) (\gamma_{\mu}D_{\mu}+m) \psi(x)$  is replaced by
\begin{eqnarray}
S_q &=& a^4 \sum_x  m \bar{\psi}(x) \psi(x) 
\nonumber \\
&+& {a^3 \over 2} \sum_{x}\sum_{k=\pm 1}^{\pm 4} 
\bar{\psi}(x) \gamma_{k} U(x,k) \psi(x+\hat{k}a)
\nonumber \\
&+& {r \over 2a} \sum_x \sum_{k=\pm 1}^{\pm 4} 
\bar{\psi}(x) U(x,k) \psi(x+\hat{k}a),
\label{quark}
\end{eqnarray}
where $\gamma_{-k}=-\gamma_{k}$ and the last term is added to avoid species doubling.
Then all the physical quantities are calculable through 
Monte Carlo (MC) simulations with importance sampling:
\begin{eqnarray}
\langle O \rangle = {\int [dU_l] \bar{O}([U_l]) e^{-S_{eff}([U_l])} 
\over \int [dU_l]e^{-S_{eff}([U_l])}} 
\approx {1 \over N_{config}} \sum_C  \bar{O}[C]. 
\end{eqnarray}
Here $C$ stands for a gluonic configuration drawn from the Boltzmann distribution.
Fermion fields must be integrated out before the simulations, which leads to $\bar{O}$ and $S_{eff}$.
Since the derivatives are approximated by finite differences, the gluonic action in Eq.(\ref{gauge})
has lattice spacing error of order $a^2$, and quark action Eq.(\ref{quark}) has error of order $a$.
To compare with the real world, the continuum limit $a \to 0$ should be eventually taken. 
On the other hand, to keep the physical volume $L^4$ unchanged, 
the number of lattice sites should be very large.
Therefore, the computational task will then be tremendously increased. 

Chinese physicists have been involved in this field since early 80's. 
Although some very interesting results were obtained,  
most of them were analytical investigations
(for review, see \cite{GuoLuo}),  due to limited computational
facilities.
Thanks to  the rapid development of high performance supercomputers in China
 in late 90's and Symanzik improvement \cite{Symanzik:1983dc} of LGT, 
this situation has changed. Today Chinese physicists play an active role 
in the lattice community \cite{Workshop}. 
Here we would like to give an overview 
of the relevant developments in recent years.

\section{Status}

\subsection{Zhongshan University}
The interests of our group cover glueballs, QCD at finite density,  supersymmetry, 
new Monte Carlo algorithms, and construction of parallel computers. 

\noindent
{\bf (a) Glueball spectrum}.
The spectroscopy of QCD in the pure gauge sector, 
i.e., the glueball masses attract considerable attention.  
In the quenched approximation,
these glueball are non $q\bar{q}$ gluonic bound states 
formed by strong self-interactions of the gluons, and their masses
vary from about 1.4 GeV to 2.5 GeV.
A lot of glueball candidates observed in BES experiments such as 
$\iota(1440)$, $f_0(1520)$, $\theta /f_J(1720)$ and $\xi(2230)$, 
produced in the  
$J/\psi$ radiative decays are within this range. 
The difficulty in experimental identification of a
glueball comes from the complexity 
in determining the quantum numbers $J^{PC}$ of these particles. 
The lattice QCD prediction for the glueball spectrum will help the experimental physicists in the their search for
glueballs. 
Concerning the lightest glueball $0^{++}$,
more accurate MC calculations 
by the IBM group
on much larger lattices, higher statistics and better algorithm gave 
$M(0^{++})\approx 1.740 \pm 0.071$ GeV, where the infinite volume  
extrapolation has been made.
Active investigation on the improvement of
the lattice techniques are still being carried out
so that the errors due to various approximations are reduced
and the glueball masses are more
accurately estimated.
We proposed an alternative way \cite{Luo:1997sa} to extract the glueball masses 
by solving the lattice QCD Schr{\"o}dinger equation. We obtained 
$M(0^{++})=1.71 \pm 0.05$  GeV, 
in nice agreement with the IBM
data. The advantage of this method, is its potential to extract the wavefunction
of a glueball. To reduce the lattice spacing errors in Eq.(\ref{gauge}) and Eq.(\ref{quark}),
we also carried out the Symanzik improvement
\cite{Luo:1996tx,Luo:1999dx,Jiang:1999hk}, which allows us to increase the precision
and perform the calculation on smaller computers.

\noindent
{\bf (b) QCD at finite density}.
This investigation is relevant for cosmology and neutron star phenomenology.  
When temperature or density is sufficiently high, a new state of matter called 
quark-gluon plasma (QGP)
is expected to form. The goal of 
Relativistic Heavy Ion Collider (RHIC) at BNL and  
Large Hadron Collider (LHC) at CERN is to create the QGP phase, and replay the evolution of the
universe. 
Although the standard lattice Lagrangian Monte Carlo method 
works very well for QCD at finite temperature, it unfortunately breaks down at finite density 
(chemical potential), because of the complex action. For SU(3) gauge theory with quarks,
the Lagrangian MC methods always lead to an unphysical critical chemical potential $\mu_c=0$ in the chiral limit. 
The Hamiltonian formulation does not have such a problem and is therefore a promising alternative.
Recently, we have developed a Hamiltonian approach 
to lattice QCD at finite density \cite{Gregory:2000pm}. 
It avoids the usual problem in the Lagrangian Monte Carlo method 
of either an incorrect continuum limit or a premature onset of the
transition to nonzero quark density as $\mu$ is raised. We
solved it in the case of free quarks and 
in the strong coupling limit. 
At zero temperature, we calculated the vacuum energy,
chiral condensate, quark number density and its susceptibility, 
as well as mass of the pseudoscalar, vector mesons and nucleons. We found that
the chiral phase transition is of first order, 
and the critical chemical potential is $\mu_C =m_{dyn}^{(0)}$ 
(dynamical quark mass at $\mu=0$). This is
consistent with $\mu_C \approx M_N^{(0)}/3$ 
(where $M_N^{(0)}$ is the nucleon mass at $\mu=0$).

\noindent {\bf (c) Dynamical quarks.}
Most conventional full QCD algorithms are expensive and do not work in the chiral limit.
The microcanonical fermionic average (MFA) method works not only on smaller computers, but also in the chiral
limit, which is very useful for studying spontaneous chiral symmetry breaking and chiral phase transition.
We generalized the MFA method to QCD \cite{Luo:1999su}. 

\noindent {\bf (d) Monte Carlo Hamiltonian.}
In  Lagrangian formulation, only the  lowest-lying state can be extracted from
the correlation function.
It is extremely difficult to compute the excited states.
We proposed a different approach: the Monte Carlo Hamiltonian method \cite{Jirari:1999jn},
designed to overcome the difficulties of the conventional approach.
The method has been well tested in quantum mechanics \cite{Jirari:1999jn,Luo:2000dz} using a regular basis in Hilbert space. 
To apply the method to many body systems and quantum field theory, stochastic basis has to be used.

\noindent {\bf (e) SUSY.}
Supersymmetry is a promising fundamental theory of elementary particles that describes a mapping 
between bosons and fermions.  It underlies many modern theories of strings and quantum gravity. 
Lattice investigations of supersymmetry may provide some new non-perturbative physics beyond the standard model.
Some numerical simulations have been done \cite{Catterall:2000rv}.

\noindent {\bf (f) Parallel computation.}
The tremendous advance in computer technology in the past decade has
made it possible to achieve the performance of a supercomputer
on a very small budget. We have built a multi-CPU cluster \cite{Cluster} 
of Pentium PCs capable of high performance parallel computations 
at a very good price/performance rate (US$\$7$/MFlops).
We believe this is the first such a machine constructed in academic institutions in China.
QCD code with an improved action has been implemented on the machine.

\noindent
\subsection{Beijing University}
Although Kogut-Susskind fermions and Wilson fermions 
have been extensively used
in numerical simulations, they are not free of problems.
There has been evidence showing
that those two approaches may give the topological charge or anomaly
incorrectly on a finite lattice.
Kaplan's domain wall fermions  
and Neuberger's overlap fermion formulation 
have attracted much attention, 
because they satisfy the Ginsparg-Wilson relations, they also produce the correct chiral modes,
anomaly and topological charge.  
For domain wall fermions there is an extra dimension and the lattice size 
in this dimension has to be very large.
Thus algorithms suitable for those new fermion approaches 
need to be developed.
Liu did some interesting work \cite{Liu:1999} in this area.

\subsection{Institute of High Energy Physics}
In quantum theory, instantons are known to be a classical solution to the $\theta$ vacuum. They might  play
an important role in confinement and chiral symmetry breaking. 
Chen, Wu and He 
calculated the glueball masses  using classical SU(2) gauge configurations \cite{Chen:2000nw}
and found \cite{Chen:2000qs} that the instantons might also give important
contributions  to the glueball spectrum. Wu is well known for his original work with Hamber on
improved fermionic actions in early 80's.

\subsection{Institute of Theoretical Physics}
Ma investigated the behavior of the gluon propagator in lattice QCD with an improved gauge action \cite{Ma:2000kn}.
 He also used an improved fermionic action to calculate the spectrum of light and heavy hadrons.

\subsection{Zhejiang University}
Ji, Li, Ying, and Zhang began  numerical simulations of LGT in early 80's. 
Due to limited computing facilities, they changed their interest to statistical physics \cite{Ying:1998nw}.
The success of  Symanzik improvement retrieves their interest in LGT.
They recently did glueball spectrum calculations on a smaller computer \cite{Zhang:1998um}.

\subsection{Beijing, Nankai and Sichuan Universities}
Chen, and Zheng and Zhu are the leading experts on  cumulant expansion of LGT and continue their
efforts in this direction \cite{Li:1997gia,Zheng:1998de,Ren:1997yq}. 
Recently, Ren, Zhu and Chen  successfully used
symbolic language to do  higher order cumulant expansion \cite{Ren:1997yq} systematically.

\section{Outlooks}

These years have seen increasing contributions by the Chinese lattice community. 
We believe they will play more active role in the near future.

\vskip 0.5cm
\noindent
{\bf Acknowledgements}\\
This work is supported by the
National Science Fund for Distinguished Young Scholars (19825117),
National Science Foundation, Guangdong Provincial Natural Science Foundation (990212) and 
Ministry of Education. 
We are grateful to S. Catterall, S. Guo, C. Huang, H. Jirari, and H. Kr\"oger for collaborations.
We also thank Prof. W.M. Suen and the organizing committee 
for their hospitality.

\end{document}